# Probabilistic assessment of the reactor vessel lifetime

V. V. Ryazanov

Institute for Nuclear Research, pr. Nauki, 47 Kiev, Ukraine, e-mail: vryazan19@gmail.com

**Highlights**
- A simple and rapid method for assessing neutron irradiation on a reactor vessel is proposed
- Queuing theory is used to estimate the lifetime of the reactor vessel walls
- The behavior of radiation defects is modeled by a random process of death and birth
- For assessments, information about neutron fluence and fast neutron energy is needed

**Abstract**

A simple and rapid method is proposed for assessing the reduction in the lifetime of steel walls of the reactor vessel under neutron irradiation. The method is based on modeling the number of radiation defects by the behavior of a general time-dependent random process of death and birth and queuing theory. Necessary data for assessments: the estimated operating time of the reactor (in years), the actual operating time of the reactor, the accumulated fluence depending on time, the temperature on the walls of the reactor vessel, the neutron absorption cross section of the steel of the reactor walls, the energy of fast neutrons striking the walls. The main problem: getting this accurate data.

## 1. Introduction

The impact of neutron irradiation on the duration of the steady state of the walls of the reactor vessel (RV) is one of the main issues in nuclear energy. The safety and reliability of the operation of a nuclear power plant are associated with the reliability of predicting changes in the viscosity characteristics of the reactor vessel material. This problem is presented in detail in many studies [1-32]; it is associated with issues of safe operation of the reactor, extending the life of the reactor, and many other tasks. The procedure usually performed when considering problems of this kind is to measure the dependence of the change in the shift of the critical brittleness temperature (or radiation embrittlement coefficient) on the temperature of the witness samples.

Thus, a large amount of data was accumulated on the radiation embrittlement of low-alloy vessel steels, on the relationship between the critical temperature of embrittlement and the resulting fluence, and on other aspects of the problem under consideration, which was then used in predicting changes in the lifetime of RV materials. Probabilistic approaches to solving this problem have been considered; recent works have used Bayesian methods, approaches related to machine learning, etc.

This article proposes a simplified probabilistic approach to solving the problem of changing the lifetime of the walls of the RV under the influence of irradiation. This is an integral approach that does not take into account many details that do not have a decisive influence on the final result. As in experimental work, the influence of neutron fluence is considered important. Lattice atoms directly displaced by incident particles are defined as a primary knocked-out atom (PKA). PVAs have an energy spectrum even in the case of monochromatic irradiation; their energy varies from zero to a certain maximum value $M_{max}$. Getting into a solid body, a fast particle is involved in a complex process of interaction with electrons and atomic nuclei in the crystal lattice. The statistical approach is based on determining the probability of the interaction process occurring. The emerging defects in the crystal lattice and crystalline crystals are considered as a statistical system, subject to the general laws of the theory of random processes, statistical physics and nonequilibrium thermodynamics. It is assumed that at some random moment this system will change its structure under the influence of irradiation. It is assumed that these changes reduce the steel time of the RV.

In almost all studies investigating the phenomenon of radiation embrittlement under the influence of defects in the crystal structure, the shift in the embrittlement temperature under the influence of irradiation is chosen as an indicator characterizing the degree of embrittlement. In this article, such an indicator is chosen to reduce the time of stable functioning of the steel structure, the life time of steel RV. This indicator can be compared with the relative narrowing of RV steel, but with the opposite sign.

The article is structured as follows. The second section describes a mathematical model of the influence of radiation effects on the lifetime of the walls of the reactor vessel. The number of radiation defects is modeled by a model of a random process of death and birth. The queuing theory is used to describe the lifetime of a RV. In the third section, the dependences of the RV lifetime on the irradiation time are calculated. The fourth section discusses the results obtained.

## 2. Mathematical model of the influence of radiation effects on the lifetime of the walls of the reactor vessel

2.1. Radiation defects.

The reaction cross section is taken as a measure of the probability density of events during the interaction of particle beams with a solid body: $\sigma = m/\varphi$, where m is the number of interactions per unit time, $\varphi$ is the flow of particles, Dimension of the interaction cross section $[\sigma]=10^{-24}см^2$. Different types of irradiation differ in the number and energy spectrum of primary knocked out atoms (PKA), which is characterized by the differential cross section $K_p(E, M)dM$, used to calculate PKA with energy M when the energy of the incident particle is E. The rate of formation of PKA with energy M is determined as follows: $P = \varphi \int_{M_d}^{M_{max}} dM K_p(E,M)$, where $\varphi$ is the flux density of incident particles. Here, only those PKAs are taken into account that have received energy exceeding the threshold energy $M_d$ required for displacement, $M_{max}$ is a certain maximum energy value.

Neutron irradiation leads to degradation of the original properties of the material. When a high-energy neutron collides with an atom in a crystal lattice, the atom is displaced or a cascade of displacements occurs in the lattice, depending on the amount of energy transferred by the neutron to the metal atom. The first atom hit by a neutron, striking other atoms, causes additional displacements in the lattice. The vacancy and its own interstitial atom formed as a result of the collision of a neutron with a lattice atom are called a Frenkel pair. As a result of the development of the cascade, volumes with a high concentration of vacancies are formed, surrounded at the periphery by zones with an increased density of interstitial atoms. In addition to displacements, large neutron fluxes, due to their energy, excite atoms and intensify their vibrations, which is accompanied by a local increase in temperature. An increase in temperature promotes radiation annealing, accompanied by the annihilation of vacancies and interstitial atoms. Further, in places where point defects accumulate, dislocation loops are formed, precipitates of elements such as copper, nickel, manganese and silicon, segregation of elements such as phosphorus and tin at interphase boundaries and grain boundaries. The basis of theoretical models of the evolution of radiation-induced defects is the kinetic equations for the concentration of point defects in a medium containing sinks. It is assumed that the concentration of radiation point defects exceeds the concentration of thermally equilibrium defects. Vacancies and interstitial atoms, migrating along the lattice, can: firstly, recombine; secondly, to form clusters of defects of the same name and, thirdly, to become sinks, which can be dislocation networks, dislocation loops, pores and other extended defects. Consequently, the rate of change in the concentration of interstitial atoms and vacancies is equal to the difference in the rates of their formation and death, which can be described by kinetic equations [32]:

$$dC_v/dt = G - RC_iC_v - C_v \sum jK_{jv},$$
$$dC_i/dt = G - RC_iC_v - C_i \sum jK_{ij}, \qquad (1)$$

where $C_v$, $C_i$ are the average concentrations of vacancies and interstitial atoms; $G=\sigma_f\varphi$ – rate of introduction of freely migrating defects ($\sigma_f$ – cross section for their formation, $\varphi$ – irradiation intensity); $R=4\pi r_{vi}(D_i+D_v)$ – recombination constant of point defects, where $r_{vi}$ – radius of mutual recombination, $D_{i,v}=D_{0i,v}\exp(-E_{mi,v}/kT)$ – diffusion coefficients of interstitial atoms and vacancies, $k$ – Boltzmann constant, $T$ – absolute temperature, $E_{m\ i,v}$ – migration energy of interstitial atom and vacancy; $K_{j\ i,v}=S_{j\ i,v}D_{i,v}$ is the coefficient of absorption of point defects by sinks of type $j$, where $S_{j\ i,v}$ are quantities characterizing the power of sinks of type $j$ for interstitial atoms and vacancies.

The embrittlement of steel walls of the RV is also influenced by impurities of phosphorus, copper and other elements, dilatation interactions between defects, the formation and evolution of partial and perfect dislocation loops, the formation of a dislocation network, the nucleation of pores and the evolution of the pore structure, radiation-induced segregation, and other reasons. Such influences can also be included in the consideration, but are not considered in this work.

### 2.2. Change in RV lifetime under irradiation

The behavior of a random process, for example, a branching random process [33], can be characterized using the generating function

$$F(s) = \sum_{k=0}^{\infty} p_k s^k, \quad |s| \leq 1, \tag{2}$$

where $p_k$ is the probability that there are k particles in the system. Let us write the generating function in the form

$$F(s) = p_0 Q(s) = Q(s)/Q = p_0/p_0(s), \; p_0 = 1/Q, \; p_0(s) = 1/Q(s), \; Q(s) = 1 + p_1 s/p_0 + ... + p_n s^n/p_0 + .... \tag{3}$$

The value $p_0=F(s=0)$ describes the probability of degeneracy of the system, that there are no particles in the system. As an example of the fulfillment of expressions (2)-(3), we can cite a large canonical ensemble [34-36].

In queuing theory (for example, [37]), the period of system occupancy, busy period, which can be compared with the lifetime of RV defects in the M/G/1 model (an exponential distribution for the flow of requests entering the system, an arbitrary distribution of the time required to service a request in the system). The average time value of busy period is equal to

$$\bar{T} = \rho/(1-\rho), \; p_0 = 1-\rho, \; \rho = 1-p_0 = 1-1/Q, \tag{4}$$

where $\rho=d\psi(\theta)/d\theta|_{\theta=0}$, the function $\psi$ characterizes the entry of elements (defects) into the system. The reactor is operating and defects are accumulating in it. This is a busy period. Then annealing can occur, - an empty period, - there are no defects. Then the reactor starts working again - a busy period.

Let us assume that the same relationships are satisfied under the influence on the system, which we denote by parameter s. The average time value of the busy period under the influence of s on the system is equal to

$$\bar{T}(s) = \rho(s)/(1-\rho(s)), \; p_0(s) = 1-\rho(s), \; \rho(s) = 1-p_0(s) = 1-1/Q(s),$$
$$\bar{T}(s)/\bar{T} = (1/p_0(s)-1)/(1/p_0-1) = (F(s)-p_0)/(1-p_0) \leq 1, \; F(s) \leq 1 \tag{5}$$

The proposed description is simplified. It is strictly valid for the large canonical ensemble [34]. But below we consider the stationary case, for which relations (2)-(5) are satisfied with a certain approximation. A consideration of the general non-stationary case using queuing theory is given in [38-40]. In [39] it is shown that the argument s from (2)-(3), (5) is written as a function of s, the argument is replaced, $s\rightarrow D(\alpha=s)=f(s)\sim\exp(-U/k_B T_e)$. The function $Q(s)$ from (2)-(5) turns out to be a function of the form $Q[D(\alpha)]$, U is the additional potential energy received by the system, $k_B$ is the Boltzmann constant, $T_e$ is the absolute temperature. Below we consider the situation when $s = e^{-U/k_B T_e}$, and $Q(s)$ where U is the energy transferred to the RV by neutrons.

### 2.3. Model of "death-and-birth"

As in [41], we use the model of death and birth [33, 42]. We assume that the incoming flow G in equations (1) enters the system one particle at a time (particle birth). A hierarchical three-level model is considered in [43]. This article assumes that the lower levels of the hierarchy are already implicitly included in the description. The concentrations of vacancies and interstitial atoms depend on time in (1) (we assume a constant input flow). Let us replace the time-dependent expressions for concentrations with average effective constant values. Taking into account concentrations after the moment when they take on stationary values leads to the same result. The generating function (2) for the process of death and birth is equal to

$$F(s, \tau, t) = \frac{1 + (1 - \xi - \eta)s}{1 - \eta s} = \sum P_n(t) z^n, \quad \rho_1(t) = \int_0^t (\mu(\tau) - \lambda(\tau)) d\tau, \tag{6}$$

$$\xi = 1 - e^{-\rho_1}/W, \quad \eta = 1 - 1/W, \quad W(\tau, t) = e^{-\rho_1}[1 + K], \quad K = \int_\tau^t e^{\rho_1(t,x)} \mu(x) dx,$$

$$\bar{n}(t) = e^{-\rho_1(t)}, \quad Var(n(t)) = e^{-\rho_1(t)}(2W - 1 - e^{-\rho_1}),$$

$$P_0(t) = \xi(t) = 1 - e^{-\rho_1(t)}/W. \tag{7}$$

Here $\bar{n}$ is the total average number of particles in the system, $Var(n(t))$ is the dispersion of the number of particles. Let us choose the parameter $\tau$ in (6)-(7) equal to 1 year in order to consider already established stationary states. It is shown in [44] that the stationary mode is established in a time of about 3 $10^7$ s, which is about 1 year. What probabilities $\lambda(t)$ and $\mu(t)$ appear in (6)-(7). In [33] says that any particle at moment t with probability $\mu(t)dt$ dies in the interval (t, t+dt) and with probability $\lambda(t)dt$ is replaced by two new particles. For the flow G entering the system of defects from (1) $\lambda(t) \sim G/n(t)$, where n(t) is the number of defects at time t. If we normalize the probability $\lambda(t)$ by dividing by $\int_\tau^{T_f} (G/n(t)) dt$, where $T_f$ is the total operating time of the reactor, we obtain that

$$\lambda(t) = \frac{G}{n(t)} \left( \int_\tau^{T_f} \frac{G}{n(t)} \right)^{-1}. \tag{8}$$

In the stationary case n(t)=const. Then in (6) $\rho_1$=0, $\lambda$=$\mu$, and assuming the constant value of G, we obtain that

$$\lambda = \mu = 1/(T_f - \tau). \tag{9}$$

The fluence is equal to $\Phi = dN/dS$, where dN is the number of particles penetrating into the sphere with cross section dS. Radiation intensity $\varphi$=dF/dS, where F=dN/dt is the flux of ionized particles. Then $\Phi = \int \varphi dt$, since $\varphi$=d(dN/dS)/dt. In expression (8) with n(t)=const, $\sigma_f[\Phi(T_f) - \Phi(\tau)] = \int_\tau^{T_f} \sigma_f \varphi dt$.

## 3. Results

Work [45] presents the dependences of the shift in the critical embrittlement temperature when witness samples are irradiated with a fluence of 17.8 $10^{18}$ n cm$^{-2}$, which corresponds to a reactor operating time of 18 years. The total operating time of the reactor is assumed to be 40 years, the value of $\tau$ is assumed to be $\tau$=1. In the practice of analyzing radiation embrittlement of materials of VVER RVs in Ukraine, the fluence of neutrons with energies above 0.5 MeV is used as a dose characteristic of fast neutrons, that is, it is assumed that neutrons with energies less than 0.5 MeV do not lead to damage to the material.

We consider the stationary case when $K = \int_\tau^t \mu(x) dx = (t - \tau)/(T_f - \tau)$. From (5)-(8) for $\Delta T = T - T(s)$ we obtain

$$\frac{\Delta T}{T} \simeq \frac{1}{1 + \frac{s}{1-s} \frac{1}{1+K}} = \frac{1}{1 + \frac{s}{1-s} \frac{T_f - \tau}{T_f + t - 2\tau}}, \quad s = e^{-U/k_B T_e}, \tag{10}$$

where U is the energy received by the walls of the reactor vessel during irradiation. The energy of the external field is added to the energy of the system as additional potential energy [46] in the grand canonical ensemble, for which relations (3), (10) are satisfied. The values vary: $T_e$ - temperature in the reactor, $\sigma_n$ - cross section for neutron absorption by the steel of the reactor walls, $E_n$ - neutron energy;

$$U=\Phi(t)\sigma_n E_n. \qquad (11)$$

For $T_f$ =40 years, $E_n$=1 mev, $\sigma_n$=1 $10^{-26}$ cm$^2$, the dependence of the fluence growth on time t, we accept a linear dependence of the form $\Phi(t)=0.98889\ 10^{18}$ t, in the interval t=0-18 years [45]. The calculated curve of the relationship between the change in the average lifetime $\Delta T=T-T(s)$, to T, where T is the average lifetime of the RV without taking into account radiation effects, T(s) - taking this into account, is shown in Fig. 1. For t=18 years, $\Delta T/T=0.631$.

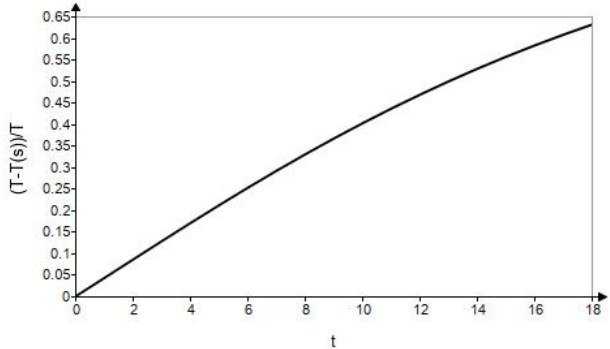

Fig.1. Dependence of the ratio of changes in the average lifetime $\Delta T=T-T(s)$, to T, where T is the average lifetime of the RV without taking into account radiation effects, T(s) – taking into account, on the reactor operating time t (in years), $T_f$=40 years, $E_n$=1 MeV, $\sigma_n$=1 $10^{-26}$ cm$^2$, $T_e$=589 K, fluence $\Phi(t)=0.98889t\ 10^{18}$ n cm$^{-2}$, in the interval t=0-18 years.

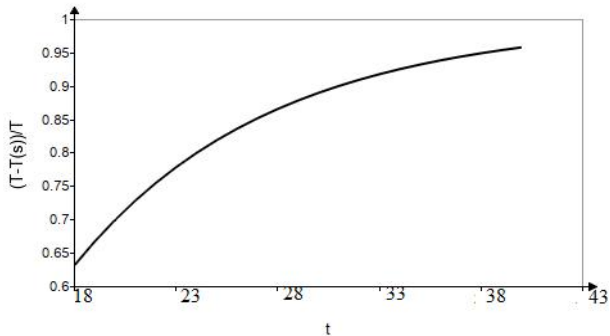

Fig.2. Dependence of the ratio of changes in the average lifetime $\Delta T=T-T(s)$, to T, where T is the average lifetime of the RV without taking into account radiation effects, T(s) – taking this into account, on the reactor operating time t (in years), $T_f$=40 years, $E_n$=1 MeV, , $\sigma_n$=1 $10^{-26}$ cm$^2$, $T_e$=589 K, fluence $\Phi(t)=(17.8+1.78t)\ 10^{18}$ n cm$^{-2}$, in the interval t=18-40 years. At t=40 years, $\Delta T/T=0.958$. A fluence of 6.3 $10^{19}$ n cm$^{-2}$ is achieved in t=43.19 years. In this case, $\Delta T/T=0.969$.

We will show the dependence on the parameters by replacing $E_n$ (in Fig. 2 it was 1 MeV, in Fig. 3 it became 0.8 MeV) and $\sigma_n$ (in Fig. 2 it was 1 $10^{-26}$ cm$^2$, in Fig. 3 it became 0.25 $10^{-26}$ cm$^2$). A significant change is visible - a decrease in the effect. So $\Delta T/T$ for t=18 years: it was in Fig. 2, $\Delta T/T=0.631$, now it is in Fig. 3, $\Delta T/T=0.196$; for t=40 years: it was in Fig. 2, $\Delta T/T=0.96$, now it is in Fig. 3, $\Delta T/T=0.556$. Changes in temperature in the reactor core also have an effect.

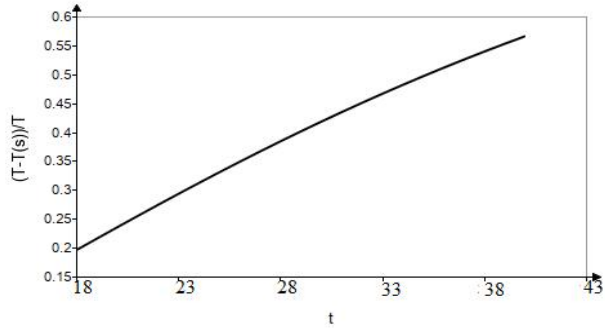

Fig. 3. Dependence of the ratio of changes in the average lifetime $\Delta T=T-T(s)$, to T, where T is the average lifetime of the RV without taking into account radiation effects, $T(s)$ – taking this into account, on the reactor operating time t (in years), $T_f=40$ years, $E_n=0.8$ MeV, $\sigma_n=0.25\ 10^{-26}$ cm$^2$, $T_e=589$ K, fluence $\Phi(t)=(17.8+1.78t)\ 10^{18}$ n cm$^{-2}$, in the interval t=18-40 years. At t=40 years, $\Delta T/T=0.566$. At t=18 years, $\Delta T/T=0.196$. A fluence of $6.3\ 10^{19}$ n cm$^{-2}$ is achieved in t=43.19 years. In this case, $\Delta T/T=0.608$.

Other sources indicate other neutron fluence values. Thus, in [47], the estimated total maximum fluence of neutrons with $E_n>0.5$ MeV at the Khmelnitsky NPP Unit No. 1, RV for the first ten fuel campaigns is $1.1\ 10^{19}$ cm$^{-2}$, with an average fluence accumulation rate of $1.1\ 10^{18}$ cm$^{-2}$ per campaign. At $T_f=40$ years and $\Phi(t=18)=19.98\ 10^{18}$ n cm$^{-2}$ it became $\Delta T/T=0.489$ in 40 years, it was 0.566 (at $\Phi(t=18)=17.8\ 10^{18}$ ), $E_n=0.8$ MeV and $\sigma_n=0.25\ 10^{-26}$ cm$^2$. At $T_f=50$ years, $\Delta T/T=0.587$, neutron absorption cross section by iron is $\sigma_n=10^{-26}$ cm$^2$; fast neutrons with $E_n>0.5$ MeV are taken into account.

If this rate of accumulation of the neutron fluence of the RV continues in the future, then the maximum permissible fluence specified in the technical safety justification for reactor plant V-320 and equal to $5.7\ 10^{19}$ n cm$^{-2}$ will be accumulated in approximately 50 years of operation, since the design service life of the VVER-1000 vessel is 40 years old.

Therefore, the task is to accurately determine such parameters as the energy of fast neutrons incident on the RV, neutron cross sections for the absorption of fast neutrons by the steels of the RV walls, fast neutron fluences, and temperature in the core. All these parameters significantly affect the change in the service life of the RV.

In [48], the relation obtained in [49] is given for the average energy of fast neutrons in a reactor of the form $\bar{E}_n(MeV)=7.42R+0.3$, where R is the ratio of the flux of neutrons with an energy greater than 3 MeV ($\Phi>3$) to the flux of neutrons with an energy greater than 1 keV ($\Phi>0.001$). In [48], measured the values of R at different points of the reactor OR, VVR [48] (left channel, level of the core center (LCC), in the same place, 20 cm below the LCC. There the same, 20 cm above the LCC. In the same place, 60 cm above LCC. Right channel of the LCC. Niche, on the central horizontal axis. In graphite reflector of the reactor, (the thickness of the graphite between the neutron detectors and the nearest fuel elements is 20-30 cm)). The average value of R is 0.153. Then $E_n=1.4357$ MeV.

The $\Delta T/T$ value calculated with this value $E_n=1.4357$ MeV and temperature on the reactor walls $T_e=290$ C is shown in Fig. 4

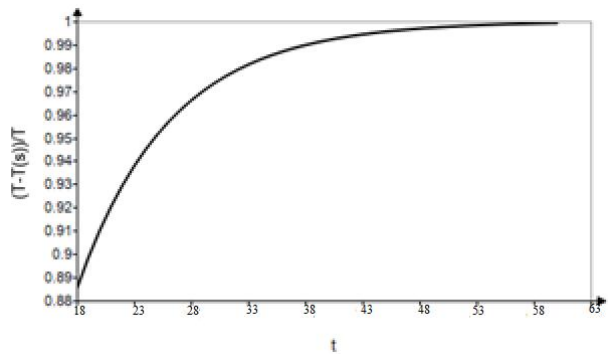

Fig.4. Dependence of the ratio of changes in the average lifetime ΔT=T-T(s), to T, where T is the average lifetime of the RV without taking into account radiation effects, T(s) – taking this into account, on the reactor operating time t (in years), $T_f$=60 years, $E_n$=1.4357 MeV, $\sigma_n$=1 $10^{-26}$ cm$^2$, $T_e$=563 K, fluence $\Phi(t)$=(19.98+1.11t) $10^{18}$ n cm$^{-2}$, in the interval t=18-60 years. At t=40 years, ΔT/T=0.992. At t=18 years, ΔT/T=0.885. At t=60 years, ΔT/T=0.999.

Since ΔT/T=1-T(s)/T, then, for example, at ΔT/T=0.992, T(s)=(1-0.992)T=0.008T, the average lifetime of the RV under the influence of irradiation is reduced for t=40 years. In Fig. 5 compared to Fig. 4 the value of $\sigma_n$ changes: $E_n$=1.4357 MeV, $\sigma_n$=0.25 $10^{-26}$ cm$^2$.

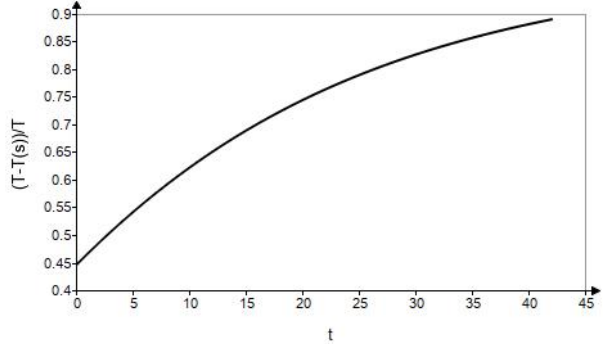

Fig. 5. Dependence of the ratio of changes in the average lifetime ΔT=T-T(s), to T, where T is the average lifetime of the RV without taking into account radiation effects, T(s) – taking this into account, on the reactor operating time t (in years), $T_f$=60 years, $E_n$=1.4 MeV, $\sigma_n$=0.25 $10^{-26}$ cm$^2$, $T_e$=563 K, fluence $\Phi(t)$=(19.98+1.11t) $10^{18}$ n cm$^{-2}$, in the interval t=18-60 years. At t=40 years, ΔT/T=0.764. At t=18 years, ΔT/T=0.446. At t=60 years, ΔT/T=0.89.

It is advisable to compare the resulting reduction in the lifetime of the RV with known results. For VVER-1000, the maximum fluence allowed is 6.3 $10^{19}$ n cm$^{-2}$. For the data in Fig. 5 this fluence will be achieved in 58.74 years. For $T_f$=60 years, $\Phi(t)$=(19.98+1.11 t) $10^{18}$ n cm$^{-2}$, $\sigma_n$=0.25 $10^{-26}$ cm$^2$, $E_n$=1.4357 MeV, t=18 years, T=563 K, it became ΔT/T=0.446, for t=40 years, ΔT/T=0.764, for t=50 years, ΔT/T=0.839; for maximum fluence $\Phi_{lim}$=6.3 $10^{19}$ n cm$^{-2}$ and duration t=58.74 years, $T_f$=60 years, ΔT/T=0.876, by $T_f$=70 years, ΔT/T=0.868. If a fluence of 1.6 $10^{20}$ n cm-2 is achieved, which is possible for VVER-440 [50], then for the data in Fig. 5 this fluence will be achieved in 145 years, with ΔT/T=0.997.

Similar results for Fig.4. But for fig. 2, ΔT/T=0.958 at t=40 years, ΔT/T=0.969 at t=43.19 years. The set of maximum fluence 6.3 $10^{19}$ n cm$^{-2}$ at $T_f$=40 years, $E_n$=1 mev, $\sigma_n$=1 $10^{-26}$ cm$^2$, $\Phi(t)$=(17.8+0.178t) $10^{18}$ n cm$^{-2}$, occurs in t=43.19 years old. For this period ΔT/T=0.969. For t=18 years, ΔT/T=0.63, for t=40 years, ΔT/T=0.958.

For Fig. 3, when $T_f$=40 years, $E_n$=0.8 MeV, $\sigma_n$=0.25 $10^{-26}$ cm$^2$, $\Phi(t)$=(17.8+1.78t) $10^{18}$ n cm$^{-2}$, time range t=18-40 years, ΔT/T=0.196, for t=18 years, ΔT/T=0.566, for t=40 years, ΔT/T=0.608, for t=43.19 years. From the comparison of Fig. 2, 4 and fig. Figure 3 shows a strong dependence of $E_n$ and $\sigma_n$ on neutron energy.

For VVER-440 the maximum fluence allowed is 3 $10^{20}$ n cm$^{-2}$ for 50 years. When $\sigma_n$=1 $10^{26}$ cm$^2$, $E_n$=1 MeV, the time of maximum fluence to reach is 252 years. In this case ΔT/T=1; for 152 years, ΔT/T=0.97. If the fluence is allowed to be 1.6 $10^{20}$ n cm$^{-2}$, then the time to reach it will be 145 years. For the parameters in Fig. 5 and for t=145 years ratio ΔT/T=0.994.

A similar characteristic of the fragility and durability of the RV metal is the relative elongation δ, shear fracture percentage. This parameter, like ΔT/T, varies from 0 to 1 (from 0 to 100%). The value of δ at which brittle fracture of a metal sample occurs can be compared with the value of T(s) at which the end of the metal's lifetime occurs. Fragility is the property of materials to collapse under the influence of external forces without residual deformation. Under significant impacts, destruction occurs at small values of δ and at large values of ΔT/T (small values of T(s)). The parameter δ is usually measured depending on the temperature, and the parameter ΔT/T is discussed above depending on the irradiation time t. But the values of these parameters correspond to each other, characterizing the end of the service life of the steel. For heavily irradiated metals, the values of the parameters δ are very small, amounting to fractions of a percent. And if, for example, δ=0.002, then T(s)=0.002T, ΔT/T=0.998. Therefore, you can give preference to Fig. 4 with higher ΔT/T values and small values of T(s).

## 4. Discussion

A simple approach to estimating the remaining lifetime of a RV after irradiation with fast neutrons is proposed. It is possible to relate this approach to the commonly used practice of measuring the critical brittleness temperature (CBT) and its shear. The simplicity of the approach can be both a disadvantage and an advantage of the method. The disadvantage is the lack of consideration of many features of the ongoing processes. But this may also be an advantage of the method, which does not take into account unimportant details, but in an integral form considers the important characteristics of the processes responsible for the main indicator of the reactor vessel - its lifetime.

The value of T(s) is not entirely clear; is it a real reduction in lifetime, or some conditional value? This value can be correlated with the change in the critical brittleness temperature. However, the latter value is also determined ambiguously. For example, an example of the spread of values was obtained within the framework of the IAEA project [51]. The corresponding results from all 8 laboratories are different.

To obtain more accurate results of the proposed approach, knowledge and strict consideration of the parameters of the problem are required. Thus, reactor steel is a complex alloy. For each component of this alloy, the neutron absorption cross sections for fast neutrons differ. Thus, for carbon, the neutron absorption cross section for fast neutrons is $\sigma_n=0.0001 \cdot 10^{-24}$ cm$^2$, and for iron - $\sigma_n=0.01 \cdot 10^{-24}$ cm$^2$. It is necessary to take into account all components of steel RV, and write $\sigma_n=\Sigma\sigma_{ni} x_i$, where $\sigma_{ni}$ is the neutron absorption cross section of fast neutrons of the i-th component, $x_i$ is the fraction of the i-th component. In a similar way, expressions for the neutron energy $E_n$ should be written, taking into account the spectrum of fast neutrons and the contribution of this spectrum to the energy of neutrons incident on the walls of the RV. More precise and strict expressions for fluence are needed.

In the general case, in the original relation for the lifetime (4), one should consider not a single system, but a network of queuing theory systems. Each network system will describe a certain section of the RV wall. This approach was used in [41]. In a simple approximation, the factor $\exp[-U/k_BT]$ includes c in the form $\exp[-U/ck_BT]$, where c is the number of systems in the queuing network. It can be shown that c~$N_0$, where $N_0$ is the number of defects at the initial time. In this work, $N_0$=1 was assumed. However, the values of c and $N_0$ can take on other values, which can be used to correct the results obtained, taking, for example, the values of $\sigma_n$ higher.

## References

1. P. G. Tipping, *Understanding and Mitigating Ageing in Nuclear Power Plants* (Elsevier Science, 2010).
2. C. K. Gupta, *Materials in Nuclear Energy Applications: Volume I* (CRC Press, 2017).
3. G. R. Odette and G. E. Lucas, Embrittlement of Nuclear Reactor Pressure Vessels, *JOM* **53**, 18 (2001).
4. J. Pu, Radiation Embrittlement, Physics **241**, Stanford University, Winter 2013.
5. S. J. Zinkle and G. S. Was, Materials Challenges in Nuclear Energy, *Acta Mater.* **61**, 735 (2013).
6. T. Dayrit, Materials Challenges in Light-Water Reactors, *Physics* **241**, Stanford University, Winter 2019.
7. S. J. Zinkle, Microstructures and Mechanical Properties of Irradiated Materials, in *Materials Issues for Generation IV Systems*, edited by V. Ghetta *et al.* (Springer, 2008).
8. O. Kutsenko, I. Kadenko, X. Pitoiset, O. Kharytonov, N. Sakhno, I. Kravchenko, Effect of neutron irradiation hardening of the base metal on the results of WWER-1000 reactor pressure vessel residual lifetime assessment, *International-journal-of-pressure-vessels-and-piping*, Volume **171**, 2019, 173-183.

9. Integrity of reactor pressure vessels in nuclear power plants: assessment of irradiation embrittlement effects in reactor pressure vessel steels, *IAEA Nuclear energy series*, No. NP-T-3.11,— Vienna : International Atomic Energy Agency, 2009.

10. E. Moslemi-Mehni, F. Khoshahval, R. Pour-Imani, M. A. Amirkhani-Dehkordi, Estimation of yield strength due to neutron irradiation in a pressure vessel of WWER-1000 reactor based on the correction of the secondary displacement model, *Nuclear Engineering and Technology*, 55, 9, 2023, Pages 3229-3240.

11. L. Zhou, J. Dai, Y. Li, X. Dai, C. Xie, L. Li, and L. Chen, Research Progress of Steels for Nuclear Reactor Pressure Vessels, *Materials* (Basel). 2022 Dec; 15(24): 8761.

12. Brumovský, M. Role of Irradiation Embrittlement in RPV Lifetime Assessment, *MRS Online Proceedings Library*, **1769**, 24 (2015). https://doi.org/10.1557/opl.2015.128.

13. C. Xu, X. Liu, Y. Li, W. Jia, Q. Quan, W. Qian, J. Yin, and X. Jin, The development of prediction model on irradiation embitterment for low Cu RPV steels, *Heliyon*, 2023; **9** (6): e16581.

14. I. Balachov, B. Fekete, D. D. Macdonald, and B. Spencer, Lifetime Estimation of a BWR Core Shroud in Terms of IGSCC, *Nuclear Engineering and Design*, **368**:110831, Nov. 2020. DOI: 10.1016/j.nucengdes.2020.110831.

15. C. Xu, X. Liu, H. Wang, Y. Li, W. Jia, W. Qian, Q. Quan, H. Zhang, F. Xue, A study of predicting irradiation-induced transition temperature shift for RPV steels with XGBoost modeling, *Nuclear Engineering and Technology*, Volume **53**, Issue 8, 2021, Pages 2610-2615, ISSN 1738-5733, https://doi.org/10.1016/j.net.2021.02.015.

16. S. Szávai, J. Dudra, Lifetime analysis of WWER Reactor Pressure Vessel Internals concerning material degradation, *20th International Conference on Structural Mechanics in Reactor Technology* (SMiRT 20) Espoo, Finland, August 9-14, 2009, SMiRT, 20-Division 2, Paper 1893.

17. Al Mazouzi, A. Alamo, D. Lidbury, D. Moinereau, S. Van Dyck, PERFORM 60: Prediction of the effects of radiation for reactor pressure vessel and in-core materials using multi-scale modelling – 60 years foreseen plant lifetime, *Nuclear Engineering and Design*, **241**, 9, 2011, pp. 3403-3415.

18. G-M. Fernandez, K. Higley, A. Tokuhiro, K. Welter, W-K. Wong, H. Yang, Status of research and development of learning-based approaches in nuclear science and engineering: A review, *Nuclear Engineering and Design*, **359**, 1, 2020, 110479.

19. T. Flaspoehlera and B. Petrovic, Radiation Damage Assessment in the Reactor Pressure Vessel of the Integral Inherently Safe Light Water Reactor (I2S-LWR), *EPJ Web of Conferences*, **106**, 03004 (2016) DOI: 10.1051/epjconf/201610603004

20. Effect of irradiation on water reactors internals, Ageing Materials Evaluation and Studies (AMES) Report No. 11., Paris (1997). Hojna A., Ernestova M., Keilova E., Kocik J., Falcnik M., Kytka M., Pesek P., Rapp M., Material characterictics of materials from Greifswald active samples/active material database/core barrel, Report NRIRez, DITI 302/419 Rev.3, (2007) (3)

21. CEA, TECNATOM and VTT, Effect of Irradiation on Water Reactors Internals, AMES report No. 11, EUR17694 EN, European Commission, Brussels–Luxemburg (1997), 4.

22. S. Chen, D. Bernard, Recommendation for computing neutron irradiation damage from evaluated nuclear data, *J. Nucl. Mater.*, **562** (2022), Article 153610

23. F. A. Garner, Radiation Damage in Austenitic Steels, (2012) In *Comprehensive Nuclear Materials*; Elsevier: Amsterdam, The Netherlands, **2012**; pp. 33–95.

24. K. Nordlund, et al. Improving atomic displacement and replacement calculations with physically realistic damage models, *Nat. Commun.*, **9** (1) (2018), pp. 1-8.

25. Nordlund, K. et al. *Primary Radiation Damage in Materials - Review of Current Understanding and Proposed New Standard Displacement Damage Model to Incorporate in Cascade Defect Production Efficiency and Mixing Effects.* NEA/NSC/DOC 9. 1−86 (Nuclear Energy Agency, OECD, Paris, 2015).

26. Wolfgang Hoffelner, *Irradiation Damage in Nuclear Power Plants, Handbook of Damage Mechanics*, pp 1427–1461, 2014.
27. Bakirov, Impact of operational loads and creep, fatigue corrosion interactions on nuclear power plant systems, structures and components (SSC), in *Understanding and mitigating ageing in nuclear power plants*, ed. by P. G. Tipping (Woodhead, Oxford, 2010), pp. 146–188
28. J. Chen, W. Hoffelner, Irradiation creep of oxide dispersion strengthened (ODS) steels for advanced nuclear applications. *J. Nucl. Mater*. **392**, 360–363 (2009)
29. Margolin B.Z., Shvetsova V.A., Gulenko A.G. Radiation embrittlement modeling in multiscale approach to brittle fracture of RPV steel. *Int. J. of Fracture*. Vol. **179** (2013), Issue 1-2, pp. 87-108.
30. Margolin B.Z., Yurchenko E.V., Morozov A.M., Pirogova N.E., Brumovsky M. Analysis of a link of embrittlement mechanisms and neutron flux effect as applied to reactor pressure vessel materials of WWER. // In: *J. Nucl.Mater*., (2013), pp. 347-356.
31. Alekseenko N.N., Amaev A.D., Gorynin I.V., Nikolaev V.A.. Radiation Damage of Nuclear Power Plant Pressure Vessel Steels //*Am. Nucl. Soc*. – La Grangeark, Illin., USA, 1997.
32. Ivanov L.I., Platov Yu.M. *Radiatsionnaia fizika metallov i ee prilozheniia [Radiation physics of metals and its applications*]. Moscow: Interkontakt, Nauka, 2002, 300 p. [in Russian].
33. T. E. Harris, *The theory of branching processes*, Shpringer-Verlag, Berlin, 1963.
34. T. L. Hill, *Statistical Mechanics: Principles and Selected Applications*, McGraw-Hill, New York (1956).
35. V. V. Ryazanov, "Simulation of the generating function for the number of particles by a branching process with immigration [in Russian]," in: *Physics of the Liquid State*, Vol. 11, Vishcha Shkola, Kiev (1983), pp. 40–44; "Modeling of thermodynamic properties of a Gibbs statistical system [in Russian]," in: *Physics of the Liquid State*, Vol. 17, Vishcha Shkola, Kiev (1989), pp. 28–41; "A constructive description of pure substances and mixtures by relations of the type of the van der Waals equation [in Russian]," in: *Physics of the Liquid State*, Vol. 18, Vishcha Shkola, Kiev (1990), pp. 5–14; "Analytic modeling of Gibbs systems [in Russian]," in: *Physics of the Liquid State*, Vol. 19, Vishcha Shkola, Kiev (1991), pp. 24–35.
36. V. V. Ryazanov, Modeling of statistical systems. I. General characteristics of the method, *Ukrainian Journal of Physics*, v. **23**, No. 6, 1978, pp. 965-972.
37. L. Kleinrock, *Queueing Systems: V. I – Theory*. New York: Wiley, 1975, Interscience. pp. 417.
38. V. V. Ryazanov, Stochastic nonequilibrium thermodynamics and time, *Ukrainian Journal of Physics*, v. **38**, No. 4, 1993, pp. 615-631.
39. V. V. Ryazanov, Stochastic modeling in nonequilibrium thermodynamics and lifetimes of systems, in: *Physics of the Liquid State*, Vol. 20, Lybid, Kiev (1992), pp. 11–36.
40. V. V. Ryazanov, Kinetics of coagulation and stochastic processes of the storage theory, In: "*Aerosols: Sciense, Indystry, Health and Environment*", 1 band, pp. 142-145. Pergamon Press, Kyoto, 1990; V. V. Ryazanov, Stochastic Description of Aerosol Systems, *Journal of aerosol Science*, 1991, Suppl.1, S59-S64.
41. V. V. Ryazanov, Influence of the external field on the semi-coagulation time and on the aerosol lifetime, *Journal of aerosol Science*, 1989, v.**20**, N8, pp. 1055-1058.
42. D. G. Kendall, On the generalized 'birth-and-death' process, *Ann. Math. Stat*. v. **19**, 1-15, 1948.
43. V. V. Ryazanov, S. Shpyrko, Hierarchic stochastic model of radiation damages lifetimes of reactor materials, The 3rd International Conference Current Problems in Nuclear Physics and Atomic Energy, June 7 - 12, 2010, Kyiv, Ukraine, The Proceedings of the Conference, pp. 538-543
44. D. A. Kornilov, V. M. Kosenkov, P. P. Silantyev, Influence of the initial dislocation structure on the kinetics of point and macroscopic defects during irradiation, *Vestn. SamU. Natural science ser*., 2016, issue 1, 69–84, [in Russian].
45. E. U. Grinik, L. I. Chirko, V. N Revka, Assessment of the radiation resource of reactor vessels of the VVER-1000 type at nuclear power plants in Ukraine. *Zbirnik naukovih prac institute of nuclear sciences*, No. 1 (14) 2005, [in Russian].

46. M. A. Leontovich. *Introduction to thermodynamics. Statistical Physics*, M.: Nauka, 1983, 416 p., [in Russian].
47. I. N. Vishnevsky, V. N. Bukanov, E. G. Vasilyeva, V. I. Gavrilyuk, A. V. Gritsenko, V. L. Demekhin, O. V. Nedelin, Monitoring the radiation load of the VVER-1000 vessel for determining its service life, International Conference of the Ukrainian Nuclear Society: Modernization of the NPP with VVER reactor, 21 - 23/09/1999, Kyiv, 1999, p.42, [in Russian].
48. B. A. Levin, Approximate methods for determining the flux of neutrons with energies greater than 0.1 MeV in experimental reactor channels, Institute of Atomic Energy. I. V. Kurchatova, Moscow, 1971, [in Russian].
49. A. Д. Kantz, Average neutron energy of reactor spectra and its influence on displacement damage, *J. Appl. Phys*. V. **34**, n. 7, p. 1944 (1963)
50. A. M. Kryukov, V. I. Lebedinsky, Irradiation embrittlement of RPV steels irradiated by high neutron fluence, *Nuclear and radiation safety*, № 1 (95)-2020, p. 2-13.
51. I. V. Orynyak, M. N. Zarazovsky, A. V. Bogdan, Method for determining the critical brittleness temperature taking into account the scatter of experimental data on impact strength, Physico-chemical mechanics of materials. – 2015. – №1. – *Physicochemical of Materials*, *Materials Science,* Consultants Bureau Plenum Publishing Corporation, (Нью-Йорк, Лондон), pp 26-36.